\documentstyle[aps,prl,psfig,twocolumn]{revtex}
\begin{document}
\wideabs{
\draft \title{ Diffusion and the Mesoscopic Hydrodynamics of Supercooled Liquids}
\author{Xiaoyu Xia\footnotemark[1] and Peter G. Wolynes\footnotemark[2]}
\address{\footnotemark[1] Department of Physics, University of Illinois, Urbana, IL, 61801\\}
\address{\footnotemark[2] Department of Chemistry and Biochemistry, University of California, San Diego, La Jolla, CA 92093}

\maketitle

\begin{abstract}
{The description of molecular motion by macroscopic hydrodynamics has a long and continuing history.  The Stokes-Einstein relation between the diffusion
coefficient of a solute and the solvent viscosity predicted using macroscopic continuum hydrodynamics, is well satisfied for liquids under ordinary to high temperature conditions, even for solutes as small as the solvent.  Diffusion in supercooled liquids near their glass transition temperature has been found to deviate by as much as 3 orders of
magnitude from that predicted by the Stokes-Einstein Relation \cite{CE96}.  Based on
the random first order transition theory \cite{KW87a,KTW89,XW00}, supercooled liquids possess a mosaic structure.  The size- and temperature-dependence of the transport anomalies
are quantitatively explained with an effective medium hydrodynamics model based on the microscopic theory of this mesoscale, mosaic structure.}
\end{abstract}

}

According to macroscopic continuum mechanics and Einstein's Brownian motion theory, the diffusion constant of a probe particle is
determined by the viscosity of the liquid through which it moves by the Stokes-Einstein Relation
\begin{equation}\label{eq:se}
D^{S-E}_t=\frac{k_BT}{4\pi \eta R},
\end{equation}
where $D^{S-E}_t$ is the translational diffusion coefficient, $\eta$ is the viscosity, and $R$ is the radius of the probe.  This relation must hold if the
solute is much larger than any relevant length scale in the liquid.  Remarkably
it is well satisfied even for self diffusion, i.e., where the radius of the solvent is the same as the solute.  It is actually not a bad relation even for
diffusion in moderately supercooled liquids ($T\geq T_g$) \cite{Rossler90,JHH80,RS65}.  Viscosity can vary by as much as 12 orders of
magnitude and diffusion coefficients vary by nearly as much.  Nevertheless there
is a noticeable discrepancy with the macroscopic hydrodynamic prediction in the
vicinity of glass transition temperature $T_g$.  In supercooled liquids around 
$T_g$, the translational diffusion is enhanced by as much as 3 orders of magnitude out of the 14 orders of magnitude change of viscosity \cite{FGSF92,CE93,CE96}!  Cicerone and Ediger have found the deviation depends strongly on the size of the probe and the temperature at which the measurement is taken \cite{CE96}.  For deviations as large as a factor of 1,000, small adjustments of hydrodynamic boundary conditions arising from the details of the probe-liquid interface will not suffice.  It is reasonable to conclude there is an additional length scale
in a supercooled liquid to explain the discrepancy.  Such a length scale arises naturally in the random first order transition theory of glasses.  In this paper
we will sketch the predictions of that theory for the deviations from the Stokes-Einstein law expected in supercooled liquids.  Ediger and others recently
have used the size-dependent deviations to infer
the size of heterogeneous domains in supercooled liquids \cite{CBE95}. In the published models, glass-forming liquids at low temperature are modeled as a two-component medium.  One component is a solid-like background in which the diffusion is slow or impossible; the other component is a liquid-like excitation in which stress can
be released quickly and the probe can move faster.  By adjusting the parameters like the lifetime, appearance rate, internal viscosity and the size of the liquid-like domains, Stillinger and Hodgdon \cite{HS93,SH94}, and Cicerone, Wagner, and Ediger \cite{CWE97} have shown it is possible to get significant deviations from the Stokes-Einstein law.  It has been appreciated \cite{CWE97} that there might not be such a clear cut distinction between two components; instead there should be a distribution of heterogeneities.  A sensible question to
ask is then how to derive this distribution from microscopic theory and
thereby to explain quantitatively the decoupling between diffusion and viscosity
in supercooled liquids.

\begin{figure}
\psfig{file=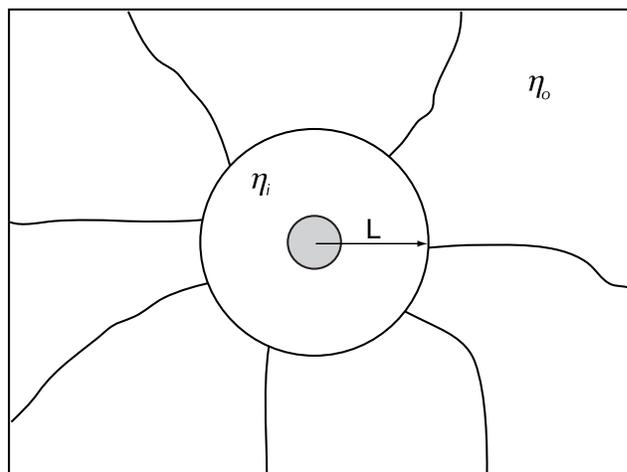,width=3.3in}
\caption
{A probe particle (shaded circle) with radius $R$ is shown in a mosaic-like
supercooled liquids.  At the instant shown, the particle is moving inside one
of the domains with size $L$.  The viscosity inside the domain is denoted
as $\eta _i$ and $\eta _o$ for the averaged outer region.  
}
\end{figure}

Recently, we developed a microscopic theory of activation barriers and fragilities in supercooled molecular liquids based on the random first order theory for the liquid-glass transition \cite{XW00}.  Several remarkable quantitative regularities emerge from theory.  The relation between the typical activation barrier in a supercooled
liquid and its configurational entropy is quantitatively obtained \cite{XW00}.  
Furthermore the theory yields the distribution of activation barriers explaining the $\beta$ parameter of the stretched
exponential relaxation functions \cite{XW00b}.  The theory also yields the size
scale of heterogeneous relaxing domains \cite{XW00,XW00b}.  We are now ready to investigate the effect of such heterogeneities on molecular hydrodynamics
in supercooled liquids.  To set the stage we briefly review the random first order theory and the microscopic theory of barriers.

The Random first order transition framework for the liquid-glass transition suggests supercooled liquids relax in an activated manner below a certain temperature, $T_A$. --The system has to overcome
some free energy barrier to reach another metastable state.  The activation barrier arises from a competition between the favorable increase in configurational entropy
(the large number of other minimum energy configurations the local region could freeze into) and an unfavorable interfacial energy (the energy cost of forming another state in the original).  This interfacial energy reflects the cost
of a domain wall-like excitation where atoms are strained and are in far from local energy minimum configurations.
Mathematically, the free energy change in forming a droplet with radius $r$ is
\begin{equation}\label{eq:mffe}
F(r)=-\frac{4}{3}\pi Ts_cr^3+4 \pi \sigma (r) r^2,
\end{equation}
where $s_c$ is the configurational entropy density driving the random first order transition.  The surface tension is strongly renormalized by the multiplicity of states which can wet the interface, much like in random
field Ising model, $\sigma(r)=\sigma _0(\frac{r_0}{r})^{1/2}$ 
\cite{XW00}, where $r_0$ is the interparticle distance.  The short range
surface tension $\sigma _0$ is nearly universal since it depends on the degree of vibrational localization in aperiodic minima.  This is usually expressed
as a universal Lindemann-like criterion for vitrification \cite{XW00}.  The free energy barrier determined by Eq.(2) can be expressed as
\begin{equation}\label{eq:fnbr}
\Delta F^{\ddagger}=\frac{3\pi\sigma^2_0r_0}{Ts_c},
\end{equation}
When the Lindemann value is substituted into the microscopic expression for $\sigma_0$ this results in an excellent description of the activation barriers in a wide range of liquids.  
The size of the droplet (or domain) is determined by the condition $F(L)=0$ giving \begin{equation}\label{eq:col}
L=(\frac{3\sigma_0r_0^{0.5}}{Ts_c})^{2/3}.
\end{equation}
This suggests regions of size $L$ are separated by domain walls and can reconfigure nearly independently.  
The configurational entropy, the driving force for reconfigurations, fluctuates around a mean value
$\overline{s_c}=\Delta \tilde{c_p}(T)\frac{T-T_K}{T_K}$, 
where $\Delta \tilde{c_p}(T)$ is the specific heat jump per unit volume at the transition,
with a magnitude $\delta s_c=\sqrt{k_B\Delta \tilde{c_p}/V^{\ddagger}}$, 
where $V^{\ddagger}=\frac{4}{3}\pi L^3$ is the volume of a typical
domain.  This results
in a corresponding variation in free energy barriers and size for each mosaic element.  This therefore gives a distribution of relaxation times and domain sizes.  This leads to a quantitative relation between the 
Kohlrausch-William-Watts exponent $\beta$ and the liquid fragility \cite{XW00b}.

The intricate mosaic structure of the supercooled liquid state presents a 
challenging problem in hydrodynamics.  We can think of each region as having 
its own viscosity $\eta _i$ determined mostly by the local relaxation time $\tau _i$.  The distribution of $\tau _i$ is determined by the entropy fluctuations.  
Of course there will be fluctuations in the higher frequency elastic modulus
of each region, $y_i$, but these should be small.  Thus $\eta _i \approx <y>\tau _i$.  If the probe is as large or larger than a typical domain it will
be surrounded several cells of the mosaic with varying $\eta _i$.  But the effect of this will be essentially the same as a homogeneous medium with the
average viscosity $\overline{\eta}=<\eta _i>$.  Only probes smaller than $L$ will be affected.  The situation should be well described by a two-zone
fluctuating viscosity hydrodynamics: a viscosity $\eta _i$ locally, which has a
fluctuating value embedded in an infinite region with the average viscosity $\overline{\eta}$.
A two-zone mean-field hydrodynamics description where the local viscosity
is fixed was proposed by Zwanzig \cite{Zwanzig89} and in a different context by Goodstein to explain ion mobility
in superfluid helium \cite{Goodstein77}.  This picture was elaborated for glasses
by Hodgdon and Stillinger \cite{HS93}.  Much of their calculation can now be taken
over, but with an additional final averaging being required.  If the viscosity has one
value $\eta_i$ in the inner zone and a different value $\eta_o$ in the outer zone, for a incompressible fluid with low Reynolds number, the fluid velocity
$v_i$ is determined by the linear Navier-Stokes equation,
\begin{equation}\label{eq:nse}
\eta \Delta v_i=\frac{\partial p}{\partial x_i},
\end{equation}
with two values of the viscosity as shown in Fig.(1).  $p$ is the pressure, and the equation of continuity,
\begin{equation}\label{eq:eoc} \bigtriangledown \cdot \vec{v}=0,
\end{equation}
must be satisfied.  For a probe particle with radius $R$ centered in such a spherical domain with size
$L$, solving Eq.(5) and (6) with the proper boundary conditions at $r=R$ and $r=L$, Hodgdon and Stillinger as well as Goodstein find,
\begin{equation}\label{eq:force}
f=4\pi \eta_oRuC_5,
\end{equation}
where $f$ is the translational drag force on the probe.  This is obtained by integrating
the pressure over the surface of the sphere, $u$ is the fluid velocity far
from the fluctuating domain, and $C_5$ is a constant determined by the ratios
of inner/outer viscosity and $L/R$.  Using slip boundary conditions, at the
surface of the probe molecule, 
\begin{eqnarray}\label{eq:c5}
C_5 & = & l[-3+3l^5+\zeta (3+2l^5)]/d_s, \nonumber \\ 
d_s & = & -2+2l^5+\zeta (4-3l+l^5+3l^6) \nonumber \\
    &   & +\zeta^2(l-1)^3(1+l)(2+l+2l^2),
\end{eqnarray}
where $l=L/R$ and $\zeta =\eta_o/\eta_i$.

From Eq.(7), we have the translational diffusion coefficient in a given domain
as
\begin{equation}\label{eq:dt1}
\tilde{D}_t=\frac{k_BT}{4\pi \eta_oRC_5},
\end{equation}
In heterogeneous supercooled liquid mosaic, the effective translational diffusion coefficient
for the whole system must be averaged over the distribution of local relaxation times,
\begin{equation}\label{eq:dt2}
D_t=\int \tilde{D}_tP(s_c)ds_c.
\end{equation}
The deviation from the homogeneous hydrodynamics results, defined as the ratio between ``real" diffusion
coefficient and the one predicted by the macroscopic Stokes-Einstein Relation, Eq.(1), is
\begin{equation}\label{eq:dvfc}
\frac{D_t}{D^{S-E}_t}=\int \frac{1}{C_5(\zeta ,l)}P(s_c)ds_c.
\end{equation}
With Eq.(3)and (4), we may express $\zeta$ and $l$ in terms of $s_c$,
\begin{equation}\label{eq:zeta}
\zeta =\frac{\eta_o}{\eta_i}=e^{\beta \overline{\Delta F}(1-\Delta F/\overline{\Delta F})}= e^{\beta \overline{\Delta F}(1-\overline{s_c}/s_c)};
\end{equation}
\begin{equation}\label{eq:lequ}
l=\frac{L}{R}=\frac{L}{\bar{L}} \frac{\bar{L}}{R}=\frac{\bar{L}}{R}(\frac{\overline{s_c}}{s_c})^{2/3}.
\end{equation}
where $\overline{\Delta F}$ and $\bar{L}$ are the typical values for
free energy barrier and domain size respectively, which can be obtained
by substituting $\overline{s_c}$ into Eq.(3) and (4).
Using a Gaussian distribution of $s_c$ with the given $\overline{s_c}$ and $\delta s_c$, we readily 
calculate the effective diffusion coefficient for a finite size probe immersed in the heterogeneous mosaic.

The experimental data are available only for a very limited number of
materials due to the difficulty of such measurements near $T_g$.  (At $T_g$ the typical
relaxation time is 1,000-10,000 seconds.) Using the results from our previous papers for the distribution of relaxation times, the size dependence of the deviation factor for a probe moving
in o-terphenyl, the most-studied glass-forming material, at $T_g$ can be calculated.  The result, which does not contain any adjustable parameters, is in excellent
agreement for small probes as shown in Fig.(2).  The deviation from the experimental data for 
large probes is to be expected for the approximated two-zone model we
applied.  It could be connected by a very modest change ($\sim 10\% $) in the local domain size and doubtlessly reflects the crudeness of the simple
geometry we have chosen.

\begin{figure}
\psfig{file=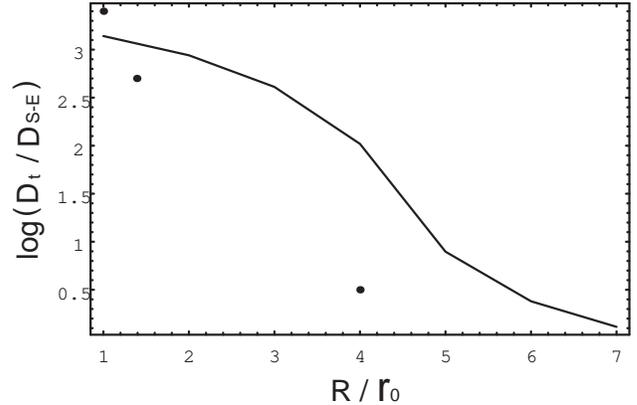,width=3.3in}
\caption
{The deviation from Stokes-Einstein Relation highly depends on the size of the probe particle as shown on the graph.  The radius of the probe particle is expressed in the unit of molecular diameters in the liquid.  For a probe with a size of several molecular distance, the deviation drops rapidly, i.e., the Stokes-Einstein Relation is retained. The points are from experiments [1] and the solid line is the prediction made by random first order theory for glass
transition.
}
\end{figure}

The temperature dependence of the deviation from the macroscopic Stokes-Einstein law is shown in Fig.(3).  The
measurement was conducted for terracene in o-terphenyl \cite{CE96}.  Again, there is a 
very good agreement between the theory and experiment over a large temperature
range.

\begin{figure}
\psfig{file=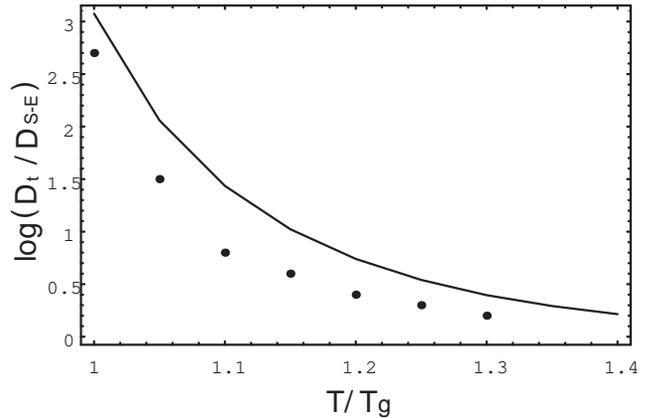,width=3.3in}
\caption
{The temperature dependence of the deviation from Stokes-Einstein Relation for terracene diffusing in supercooled o-terphenylis shown.  The theoretical results
(solid line) agrees excellently with experimental data (points) [1].
}
\end{figure}

The deviation factors depend on the fragility, $D$, of the
liquids if they are all measured at the glass transition temperature $T_g$.  The degree of the heterogeneity, as measured by $\delta s_c/\overline{s_c}$, decreases as the liquid gets stronger \cite{XW00b}.  For very strong glass-forming liquids (like $\mathrm{SiO_2}$ with $D=150$), one would
expect the macroscopic Stokes-Einstein Relation to hold well.  The theory predicts large deviations only for very fragile liquids (with $D$
approximately smaller than 20) as shown in Fig.(4).  The plot is for self-diffusion (or probe with
similar size with the liquid molecule).  Unfortunately presently, most
of the experiments have been reported only for a very narrow range of
substances with $D\sim 10$.  We present the fragility dependence thus as a prediction to encourage investigations over a wider range of liquids.

\begin{figure}
\psfig{file=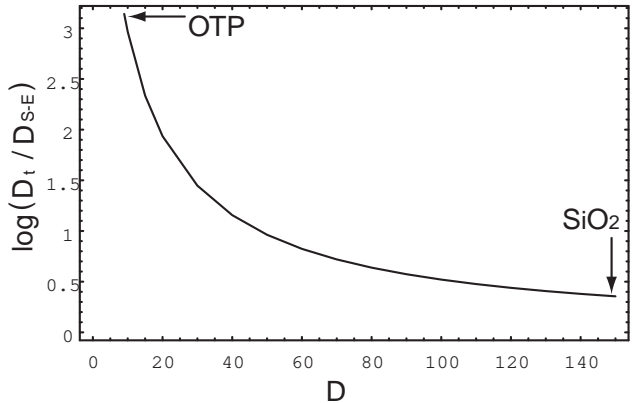,width=3.3in}
\caption
{The fragility of the supercooled liquids has important effect on the observed
Stokes-Einstein Relation deviation.  For fragile liquids like o-terphenyl, such
deviation is as large as 3 orders of magnitude, while in strong liquids like
$\mathrm{SiO_2}$, it should be only minor.  The plot is for self-diffusion at glass
transition temperature $T_g$.
}
\end{figure}

There are a couple of important points to be made clear before we conclude our remarks.
One is that we use the ``average" or the measured viscosity as the outer viscosity $\eta _o$ in the calculation.  A more thorough theory should treat
simultaneously
the fluctuations inside and outside the domain.  Nevertheless the approximation to separate these two terms turns out to be a sound one.  The physics underlying the simplification is the asymmetry in the contribution to the diffusion from the ``fast" regions and ``slow" regions.
From the structure of Eq.(10), it is clear fast regions with relatively smaller
local relaxation times (or viscosities) have much larger contributions to the averaged
effective diffusion coefficient.  A little algebra shows the largest contribution
is from regions with $s_c \sim 0.67\overline{s_c}$.  For these regions the local
relaxation time is more than 5 orders of magnitude faster than the measured relaxation
time (about 1,000-10,000 seconds).  For practical purposes, the outer
part of the system acts homogeneously for these fast regions.

The other comment concerns the treatment of smaller domains.  As we have shown, at $T_g$, the typical domain size is about 7 times of the molecular
distance.  From Eq.(4), we know fast regions are also larger compared to the
average.  For the same reason we discussed above, these fast (or larger) regions
have a much greater contribution to the calculated diffusion.  The two-zone
approximation we use will hold well.  However, for probes with very large
size comparable to the typical domain size, such an approximation may pose
a problem as the diffusion of rubrene in o-terphenyl indicates.  When the probe
particle occupies multiple domains, an additional average over these domains
should be taken to get more accurate results.  Nevertheless, the model developed
in this paper should serve as a good starting point even for larger probe
diffusion.

In summary, the mosaic picture of the supercooled liquids
based on the random first order transition theory explains the unusual decoupling of translational diffusion from the macroscopic hydrodynamics of glass-forming liquids.  The picture has as an essential element the new length scale
of dynamically correlated regions.  This length is determined to a good
approximation by a microscopic theory of the glassy state itself.  The mosaic
picture also has interesting consequences for the frequency dependence
and nonlinearity of the response of a translating probe.  Similarly the mesoscale correlation will affect rotational hydrodynamics and relative motion
of probes in the liquid as is important for diffusion controlled reactions.  We leave these problems for the future. 

This work is supported by NSF grant CHE-9530680.  We are happy to dedicate this work to Bruce Berne a pioneer
of not only computer simulation but also of the hydrodynamic view of molecular motion in
the modern era.

\end{document}